\documentclass[twocolumn,RNAAS]{aastex631}
\usepackage{aas_macros} 
\usepackage{natbib} 
\shorttitle{Millisecond Pulsars}
\shortauthors{Halder et al.}
\newcommand{\bed}{\begin{displaymath}}
\newcommand{\eed}{\end{displaymath}}
\newcommand{\bei}{\begin{itemize}}
\newcommand{\eei}{\end{itemize}}
\newcommand{\bef}{\begin{figure}}
\newcommand{\eef}{\end{figure}}
\newcommand{\ben}{\begin{enumerate}}
\newcommand{\een}{\end{enumerate}}
\newcommand{\beq}{\begin{equation}}
\newcommand{\eeq}{\end{equation}}
\newcommand{\ber}{\begin{eqnarray}}
\newcommand{\eer}{\end{eqnarray}}

\newcommand{\pdot}{\mbox{$\dot P$}}

\newcommand{\lsim}{\raisebox{-0.3ex}{\mbox{$\stackrel{<}{_\sim} \,$}}}

\begin{document}

\title{Defining Millisecond Pulsars}

\author{Priyam Halder}
\affiliation{SASTRA University, Thanjavur 613401, India}

\author{Satyaki Goswami}
\affiliation{IIT Madras, Chennai 600036, India}

\author{Protyusha Halder}
\affiliation{Lovely Professional University, Jalandhar 144411, India}

\author{Uday Ghosh}
\affiliation{ISI-North-East Centre, Tezpur 784501, India} 

\author[0000-0003-2231-6658]{Sushan Konar}
\affiliation{NCRA-TIFR, SP Pune University Campus, Pune 411007, India}

\correspondingauthor{Sushan Konar}
\email{sushan.konar@gmail.com}

\begin{abstract}
Millisecond  pulsars  (MSP)  are  an important  subclass  of  rotation
powered  pulsars (RPP),  traditionally defined  as those  with $P_s  <
20-30$~ms and  $B_s \lsim 10^{10}$~G. We re-examine this  definition by
applying Gaussian  mixture model (GMM) analysis  to identify distinct
clusters within the RPP population and find that the MSPs appear to be
better demarcated by the condition $\mathbf{P_s \lsim 16}$~ms.
\end{abstract}

\section{}

Approximately  4000  neutron  stars  have been  observed  since  their
serendipitous  discovery more  than  fifty years  ago~\citep{hewis68}.
Despite  widely  divergent  observational  characteristics  and  their
consequent  grouping  in  a  somewhat large  number  of  observational
classes, it  is possible  to classify these  neutron stars  into three
main types depending on their mode of energy generation.  Accordingly,
they can either be {\bf a)} rotation powered pulsars (RPP), or {\bf b)}
accretion  powered  pulsars  or   {\bf  c)}  internal  energy  powered
objects~\citep{konar13,konar16c,konar17e}.

However, most  of the observed neutron  stars belong to the  RPP class
and  are powered  by the  loss of  rotational energy  due to  magnetic
braking. The basic observed quantities of a pulsar are its spin-period
($P_s$)  and the  period derivative  (\pdot), and  the most  important
derived  quantity is  the dipolar  component of  the surface  magnetic
field given by \citep{manch77} -
\beq
B_{\rm s}
\simeq 3.2 \times 10^{19}
\left(P_{\rm s}/s\right)^{\frac{1}{2}}
\left(\pdot/ss^{-1}\right)^{\frac{1}{2}}~{\rm G}\,.
\label{eq01}
\eeq
%

An  important  sub-class   of  RPPs  is  the   rapidly  spinning  {\em
  Millisecond Pulsars} (MSP). Starting  with the discovery of B1937+21
($P_s$  = 1.558~ms)  by \cite{backe82},  initially there  were only  a
handful   of  pulsars   with   $P_s  <   10$~ms   which  were   called
MSPs~\citep{bhatt91a}.  As the number of ultra-fast pulsars increased,
MSPs   came   to   be   known   as    those   with   $P_s   <   20   -
30$~ms~\citep{corde97,lorim08},  a group  that currently  accounts for
$\sim15$\% of  the Galactic RPPs.   These MSPs, recycled  pulsars with
both  ultra-short spin-periods  and weak  magnetic fields  ($B_s \lsim
10^{10}$~G), are understood to  have different evolutionary histories.
The  two  characteristic  properties,  namely  $P_s$  and  $B_s$,  get
inter-related through the process of  MSP formation either by {\bf \em
  recycling}   of   ordinary   radio    pulsars   or   by   {\bf   \em
  accretion-induced  collapse}  of  white  dwarfs,  during  long-lived
accretion  phases  of the  low-mass  or  the intermediate  mass  X-ray
binaries (LMXB, IMXB)~\citep{radha82,bhatt91a,tauri00}.

The direct support  for MSP formation in X-ray binaries  came from the
discovery   of   coherent   millisecond  X-ray   pulsations   in   SAX
J1808.4~\citep{chakr98,wijna98}  and  subsequently   in  many  similar
objects~\citep{wijna05,watts12,  patru21}.   On  the other  hand,  the
Galactic centre gamma-ray  excess appears to have  an easy explanation
if       the       MSPs        there       have       formed   due to 
accretion-induced-collapse~\citep{gauta22}.

\bef
\includegraphics[width=8.5cm]{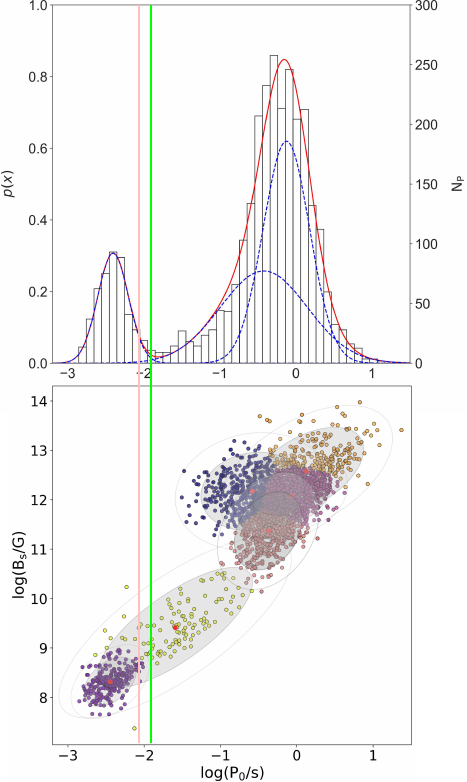}
\caption{Top :  1D GMM decomposition of  the $P_s$ data.  Bottom  : 2D
  GMM  decomposition of  the combined  $P_s, B_s$  data. The  ellipses
  denote  the   2$\sigma$  and  3$\sigma$  contours   of  the  cluster
  components.  The pink  and the green lines show the  $P_s$ maxima of
  the corresponding contours of the MSP cluster. [Data - ATNF Pulsar
  catalog]}
\label{f_gmm}
\eef

In order to  understand these different formation  scenarios and their
influences on the  characteristics properties of neutron  stars, it is
important to  study the population  of MSPs, requiring  a well-defined
criterion for identification of MSPs.   So far the  classification of
MSPs has depended upon the fact that the $P_s$ data showed a clear gap
around $P_s \sim 20 - 30$~ms in the beginning.  With a much larger RPP
population, this demarcation is not so obvious any more.

But defining MSPs with accuracy is  not easy and a rigorous definition
must  involve both  $P_s$  and $\pdot_s$  (or,  equivalently $P_s$  \&
$B_s$) since these parameters get intertwined during MSP formation. An
attempt to  do so  was made by  \cite{story07}, assuming  the Galactic
disc MSPs  to form through  direct recycling in LMXBs.   Clearly, this
has  limited applicability.   For example,  in Globular  Clusters, the
MSPs form  through complex  routes of  binary evolution,  resulting in
somewhat different distribution  of $P_s$ and $B_s$  compared to those
in the disc~\citep{konar10,konar19a}.

In this  work, we attempt to  arrive at a semi-rigorous  definition of
MSPs, using  both the $P_s$  and $B_s$  data. An earlier  GMM analysis
showed that the  $P_s$ data is best  fit by three components  - a) the
MSPs, b) the  pulsars recycled in high-mass X-ray binaries  and c) the
ordinary  radio  pulsars~\citep{konar17e}, though this  work  did  not
attempt to quantitatively demarcate individual Gaussian components. In
the current work - {\bf a)} we  redo the 1D GMM as the current data has
much larger   size  and find  a {\em   qualitatively}   similar result
(Fig.\ref{f_gmm}, top); {\bf b)} - we perform a 2D GMM analysis on the
combined $P_s,  B_s$ set.   The 2D analysis  brings out  6 independent
Gaussian clusters as shown in Fig.\ref{f_gmm} (bottom), similar to the
results     obtained     by     a    recent     work     on     pulsar
classification~\citep{reddy22}.

Ideally, a 2D analysis should be more reliable as it includes both the
characteristic parameters.  Unfortunately, the  2D case suffers from a
serious drawback due to the unavailability of $B_s$ values for a large
number of RPPs. Currently,  ATNF~\citep{manch05a} has $P_s$ values for
3295 objects, but $B_s$ values for only 2514 of them.  This is why, it
is essential to  combine the results from both 1D  and 2D GMM analysis
to arrive at a `good' definition of MSPs.

The $P_s$ values  obtained from 1D GMM at 95\%  (2$\sigma$) and 99.7\%
(3$\sigma$)  population  cut-off  are 10~ms  and  16~ms  respectively;
whereas  those obtained  from 2D  GMM are  8~ms (2$\sigma$)  and 12~ms
(3$\sigma$). Clearly, the 2D GMM limits are well within those obtained
from 1D GMM.  On the other  hand, 95\% cutoff for the middle component
of 1D GMM is 27~ms which allows  for a robust upper cutoff for MSPs at
16~ms.   Therefore, even  keeping  in  mind that  GMM  analysis is  an
unsupervised machine learning algorithm which  does not allow for very
high accuracy, it would perhaps not be too unreasonable to define MSPs
as the {\bf \em subclass of rotation powered pulsars with $\mathbf{P_s
    \lsim 16}$~ms}.

\begin{acknowledgments}
  Excellent hospitality at MANUU, Hyderabad during ADAP'23, organised
  by Priya \& Najam Hasan, allowed us to finally settle this question.
\end{acknowledgments}

\bibliography{adsrefs}
\bibliographystyle{aasjournal}

\end{document}